# Diffractive Interconnects: All-Optical Permutation Operation Using Diffractive Networks

Deniz Mengu[1,2,3], Yifan Zhao[1,3], Anika Tabassum[1,3], Mona Jarrahi[1,3], Aydogan Ozcan[1,2,3,4,*]

[1] Electrical and Computer Engineering Department, University of California, Los Angeles, CA, 90095, USA
[2] Bioengineering Department, University of California, Los Angeles, CA, 90095, USA
[3] California NanoSystems Institute, University of California, Los Angeles, CA, 90095, USA
[4] Department of Surgery, David Geffen School of Medicine, University of California, Los Angeles, CA, 90095, USA.
* Corresponding author: ozcan@ucla.edu

**Abstract:** Permutation matrices form an important computational building block frequently used in various fields including e.g., communications, information security and data processing. Optical implementation of permutation operators with relatively large number of input-output interconnections based on power-efficient, fast, and compact platforms is highly desirable. Here, we present diffractive optical networks engineered through deep learning to all-optically perform permutation operations that can scale to hundreds of thousands of interconnections between an input and an output field-of-view using passive transmissive layers that are individually structured at the wavelength scale. Our findings indicate that the capacity of the diffractive optical network in approximating a given permutation operation increases proportional to the number of diffractive layers and trainable transmission elements in the system. Such deeper diffractive network designs can pose practical challenges in terms of physical alignment and output diffraction efficiency of the system. We addressed these challenges by designing misalignment tolerant diffractive designs that can all-optically perform arbitrarily-selected permutation operations, and experimentally demonstrated, for the first time, a diffractive permutation network that operates at THz part of the spectrum. Diffractive permutation networks might find various applications in e.g., security, image encryption and data processing, along with telecommunications; especially with the carrier frequencies in wireless communications approaching THz-bands, the presented diffractive permutation networks can potentially serve as channel routing and interconnection panels in wireless networks.

# 1 Introduction

Permutation is one of the basic computational operations that has played a key role in numerous areas of engineering e.g., computing[1], communications[2], encryption[3], data storage[4], remote sensing[5] and data processing[6]. Historically, electronic integrated circuits have been the established implementation medium for the permutation operation and other space-variant linear transformations, while the research on optical computing has been mainly focused on using the Fourier transform approximation of thin lenses covering various applications in space-invariant transformations e.g., convolution/correlation. On the other hand, as photonic switching devices and optical waveguide technology have become the mainstream communication tools on high-end applications e.g., fiber optic networks, supercomputers and data centers, various approaches have been developed towards all-optical implementation of permutation operation and other space-variant transformations based on e.g., Mach-Zehnder interferometers[7], optical switches[8], photonic crystals[9], holographically recorded optical elements[10–12], off-axis lenslet arrays[13,14] and arrays of periodic grating-microlens doublets[15]. The development of compact, low-power optical permutation and interconnection devices can have significant impact on next-generation communication systems e.g., 6G networks[16,17], as well as other applications such as optical data storage[18] and image encrypting cameras[19–21].

With the widespread availability of high-end graphics processing units (GPU) and the massively growing amounts of data, the past decade has witnessed major advances in deep learning, dominating the field of digital information processing for various engineering applications including e.g., image segmentation and classification[22–25], natural language processing[26,27], among others[28]. The statistical inference and function approximation capabilities of deep neural networks have also been exploited to produce state-of-the-art performance for computational inverse problems in many imaging and sensing applications including e.g., microscopy[29–35], quantitative phase imaging[36–41] and others[42–51]. Beyond these data processing tasks, deep learning can also provide task-specific solutions to challenging inverse optical design problems for numerous applications including nanophotonics[52,53], metamaterials[54], imaging and sensing[55–60]. However, as the success and the applications of deep learning grow further, the electronic parallel computing platforms e.g., GPUs, hosting deep neural networks and other machine learning algorithms have started to bring some limitations due to their power- and bandwidth-hungry operation. Moreover, the pace of the advances in computational capacity of the integrated circuits has fallen behind the exponential increase predicted by the Moore's law[61]. These factors have fueled a tremendous amount of effort towards the development of optical machine learning schemes and other photonic computing devices that can partially reduce the computational burden on the electronics leading to power-efficient, massively parallel, high-speed machine learning systems. While most of the arising optical computing techniques rely on integrated photonic devices and systems compatible with the integrated waveguide technology[62–68], an alternative option towards exploiting photons for machine learning and the related computing tasks is to use complex modulation media and free-space light propagation and diffraction, which is particularly suitable for visual computing applications where the information is already carried by the optical waves (e.g., of a scene or target object) in free-space[69].

Motivated by these pressing needs, Diffractive Deep Neural Networks ($D^2NN$)[70,71] have emerged as an optical machine learning framework that utilizes deep learning to engineer light-matter interactions over a series of diffractive surfaces so that a desired statistical inference or deterministic computing task is realized all-optically as the light waves propagate through structured surfaces. According to this framework, the physical parameters determining the phase and/or amplitude of light over each independently controllable unit, i.e., the 'diffractive neuron', are updated through the stochastic gradient descent and error-backpropagation algorithms based on a loss function tailored specifically for a given task. The weights of the connections between the diffractive neurons/features on successive layers, on the other hand, are dictated by the light diffraction in free-space. Once the deep learning-based training, which is a one-time effort, is completed using a computer, the resulting transmissive/reflective diffractive layers are fabricated using e.g., lithography or 3D printing, to physically form the diffractive network that *completes* a given inference or computational task at the speed of light using entirely passive modulation surfaces, offering a task-specific, power-efficient and fast optical machine learning platform.

Based on the $D^2NN$ framework, here we demonstrate diffractive optical network designs that were trained to all-optically perform a given permutation operation between the optical intensities at the input and output fields-of-view, capable of handling hundreds of thousands of interconnections with diffraction limited resolution. We quantified the success of the presented diffractive optical networks in approximating a given, randomly-selected permutation operation as a function of the number of diffractive neurons and transmissive layers used in the diffractive network design. We also laid the foundations toward practical implementations of diffractive permutation networks by investigating the impact of various physical error sources, e.g., lateral and axial misalignments and unwanted in-plane layer rotations, on the quality/accuracy of the optically-realized interconnection weights and the permutation operation. Moreover, we showed that the diffractive optical permutation networks can be trained to be resilient against possible misalignments as well as imperfections in the diffractive layer fabrication and assembly. Finally, we present the first proof-of-concept experimental demonstration of diffractive permutation networks by all-optically achieving a permutation matrix of size 25×25, effectively realizing 625 interconnections based on 3D-printed diffractive layers operating at the THz part of the spectrum.

The presented diffractive optical permutation networks can readily find applications in THz-band communication systems serving as communication channel patch panels; furthermore, the underlying methods and design principles can be broadly extended to operate at other parts of the electromagnetic spectrum, including the visible and IR wavelengths, by scaling each diffractive feature size proportional to the wavelength of light[72] and can be used for image encryption in security cameras[73] and optical data storage systems, among other applications[74–77].

## 2 Results

Figure 1 illustrates the presented free-space permutation interconnect concept designed around diffractive optical networks using the D$^2$NN framework. As shown in Fig. 1, the presented permutation interconnect scheme does not use any standard optical components such as lenses, and instead relies on a series of passive, phase-only diffractive surfaces. Due to the passive nature of these layers, the diffractive optical network shown in Fig. 1 does not consume any power except for the illumination light, providing a power-efficient permutation operation in a compact footprint of ~600λ along the longitudinal axis), which could be further squeezed as needed. The 5-layer diffractive optical permutation network design shown in Fig. 1 was trained through supervised learning to all-optically realize a desired permutation operation, $P$, between the light intensity signals at the input and output FOVs, each with $N_i = N_o = 400$ (20×20) individual pixels of size 2λ×2λ. Stated differently, this permutation operation controls in total of $N_i N_o = 0.16$ million optical intensity connections.

The supervised nature of the training process of the diffractive permutation network necessitates the use of a set of input-output signal pairs (examples that satisfy $P$) to compute a penalty term and the associated gradient-based updates with respect to the physical parameters of each diffractive neuron at every iteration. We set the optical signal of interest at the input and the output of the diffractive permutation scheme to be the light intensity, and as a result, the deep learning-based evolution of the presented diffractive permutation network shown in Fig. 1 was driven based on the mean-squared error (MSE) (see Methods section) between the ground-truth and the all-optically synthesized output intensity patterns at a given iteration. Since this loss function acts only on the light intensity, the diffractive optical network can enjoy an output phase-freedom in synthesizing the corresponding transformed optical intensity patterns within the output field-of-view. The light intensity, $I$, is related to the complex-valued field, $U$, through a nonlinear operation, $I = |U|^2$. If a pair of input-output complex-fields exists for a given diffractive network, i.e., $\{U_{in}, U_{out}\}$ and $\{U'_{in}, U'_{out}\}$, then the input field $U''_{in} = \alpha U_{in} + \beta U'_{in}$ will create the ideal output field $U''_{out} = \alpha U_{out} + \beta U'_{out}$ at the output plane. In terms of the associated intensities, however, this direct linear extension does not hold since $|\alpha U_{out}|^2 + |\beta U'_{out}|^2 \neq |U''_{out}|^2$, making it challenging (in terms of the data generalization capability) to design diffractive optical networks for achieving a general purpose intensity-to-intensity transformation such as a permutation operation. To overcome this generalization challenge, we trained our diffractive permutation networks using ~4.7 million randomly generated input/output intensity patterns that satisfy the desired $P$, instead of a standard benchmark image dataset (see the Methods).

After the training phase, we blindly tested each diffractive permutation network with test inputs that were never used during the training. Figure 2 illustrates 6 different randomly generated blind testing inputs along with the corresponding all-optically permuted output light intensities. In the first two randomly generated input patterns shown in Fig. 2A, there is light coming out of all the input pixels/apertures at different levels of intensity. In the next two test input patterns shown in Fig. 2A, on the other hand, nearly half of the input apertures have nonzero light intensity and finally, the last two test inputs contain only 10 and 11 pixels/apertures with light propagating towards the 1$^{st}$ layer of the diffractive permutation network. When tested on 20K randomly generated blind testing input intensity patterns with different sparsity levels, the 5-layer diffractive permutation network shown in Fig. 1 achieves 18.61 dB peak-signal-to-noise ratio (PSNR), very well matching the ideal output response. For this randomly generated 20K testing data, Figure 2B also shows the distribution of PSNR as a function of the number of input pixels with nonzero light intensity, which reveals that the diffractive permutation network can permute relatively sparser inputs with a higher output image quality, achieving a PSNR of 25.82 dB.

In addition to randomly generated blind testing inputs, we further tested the diffractive permutation network shown in Fig. 1 on 18.75K EMNIST images; note that this diffractive network was trained only using randomly generated input/output intensity patterns that satisfy $P$ and the EMNIST images constitute not only blind testing set but also a significant deviation from the statistical distribution of the training images. The input field-of-view contains the permuted EMNIST images ($P^{-1}$) and the diffractive network inverts that permutation by all-optically performing $P$ to recover the original images at the output plane (see Fig. 2). The performance of the diffractive permutation network was quantified based on both PSNR and Structural Similarity Index Measure (SSIM). With $N_L$=40K diffractive neurons on each layer, the 5-layer diffractive permutation network shown in Fig. 1 provides 19.18 dB and 0.85 for PSNR and SSIM metrics, respectively, demonstrating the generalization capability of the diffractive network to new types of input image data never seen during the training phase.

## 2.1 Impact of the number of diffractive layers and features

Next, we investigate the performance of diffractive permutation networks as a function of the number of diffractive neurons trained within the system. Towards this goal, in addition to the 5-layer design shown in Fig. 1 that has in total of $N$=200K trainable diffractive features, we trained diffractive permutation networks consisting of 4, 3 and 2 modulation surfaces. The physical design parameters such as the size/width of the diffractive surfaces, layer-to-layer distances and the extent of the input and output fields-of-view, were kept identical to the ones in the 5-layer network design. In other words, these new diffractive networks are designed and trained exactly in the same way as the previous 5-layer network except they contain fewer diffractive layers. Figures 3A and 3B provide a quantitative comparison between these 4 diffractive permutation networks. While Fig. 3A illustrates the mean PSNR and SSIM values achieved by each diffractive network for recovering EMNIST images, Fig. 3B demonstrates the mean-squared-error (MSE) between the desired permutation operation and its optically realized version ($P_{D^2NN}$) as a function of the number of diffractive layers utilized in these designs (see Supplementary Fig. S1 and the Methods section on the estimation of $P_{D^2NN}$). According to the permutation operator error shown in Fig. 3B, the performance improvement of the system increases drastically with the additional diffractive layers up to the 4-layer design that represents a critical point in the sense that the inclusion of a $5^{th}$ diffractive surface brings a relatively small improvement. The reason behind this behavior is the fact that the number of diffractive features, $N$, in the 4-layer diffractive permutation network matches the space-bandwidth product set by our input and output FOVs, i.e., $N_iN_o = 400{\times}400$=160K. In other words, Fig. 3B reveals that when the number of phase-only diffractive modulation units $N$ matches or exceeds $N_iN_o$, the diffractive optical network can achieve a given linear transformation between the input and output intensities with a very low error, i.e., $P_{D^2NN} \approx P$; for example, the MSE between $P_{D^2NN}$ and $P$ in the case of a 4-layer design was found to be $6.63 \times 10^{-5}$. For $N < N_iN_o$, the error between $P_{D^2NN}$ and $P$ increases accordingly, as shown in Fig. 3B.

The benefit of having $N \geq N_iN_o$ is further revealed in the increased generalization capability of the diffractive network as shown in Fig. 3A. Since the EMNIST images were not used during the training, they represent completely new types of input intensity patterns for the presented diffractive optical networks. The SSIM (PSNR) values achieved by the 4-layer diffractive network is found as 0.75 (16.41 dB) for the optical recovery of the permuted EMNIST images. These numbers are significantly higher compared to the performance of the 3-layer and 2-layer diffractive designs that can attain SSIM (PSNR) values of 0.46 (12.91 dB) and 0.30 (12.08 dB) for the same task; furthermore, the 5-layer diffractive network design shown in Fig. 1 outperforms the others by achieving 0.85 (19.18 dB) for the same performance metrics. The visual comparison of the input-output intensity patterns depicted in Fig. 3C further supports this conclusion, where the noise due to the crosstalk between interconnection channels decreases proportional to the number of diffractive layers in the system.

## 2.2 Vaccination of diffractive permutation networks

With sufficiently large number of phase-only diffractive neurons/features, the diffractive networks can optically realize permutation operations with e.g., 0.16 million channels between the input and output pixels as shown in Fig. 3. In fact, the number of interconnects that can be optically implemented through diffractive networks can go far beyond 0.16 million, given that the size/width of the diffractive surfaces and the number of diffractive layers can be increased further depending on the fabrication technology and the optomechanical constraints of the system. In addition, as the number of diffractive layers increases in a diffractive network architecture, their forward model can better generalize to new, unseen data as shown in Fig. 3A.

On the other hand, deeper diffractive optical network designs are more susceptible to misalignments that are caused by the limitations of the optomechanical assembly and/or the fabrication technology that is utilized. It was shown that diffractive optical networks trained for statistical inference tasks e.g., all-optical object classification, can be vaccinated against misalignments and other physical error sources, when the factors creating these nonideal conditions were incorporated into the training forward model, which was termed as vaccinated-D²NNs or v-D²NNs[78]. Specifically, v-D²NN expands on the original D²NN framework by modeling possible error sources as random variables and integrating them as part of the training model so that the deep learning-based evolution of the diffractive surfaces is guided towards solutions that are resilient to nonideal physical conditions and/or fabrication errors. Towards practical applications of diffractive permutation networks, we quantified the impact of optomechanical errors and applied the v-D²NN framework to devise robust solutions that can achieve a given interconnect operation despite fabrication tolerances.

In our numerical study depicted in Fig. 4, we considered 4 different misalignment components representing the 3D misalignment vector of the $l^{th}$ diffractive layer, $\left(D_x{}^l, D_y{}^l, D_z{}^l\right)$ and their in-plane rotation around the optical axis denoted as $D_\theta{}^l$. Each of these 4 misalignment components were defined as independent, uniformly distributed random variables, $D_*{}^l \sim U(-\Delta_*, \Delta_*)$, with $\Delta_*$ defined as a function of a common auxiliary parameter, $v$. The lateral misalignments parameters, $\Delta_x$ and $\Delta_y$, determining the range of $D_x{}^l$ and $D_y{}^l$, respectively, were set to be $0.67\lambda v$, i.e., $D_x{}^l \sim U(-0.67\lambda v, 0.67\lambda v)$ and $D_y{}^l \sim U(-0.67\lambda v, 0.67\lambda v)$, where $\lambda$ denotes the wavelength of the illumination light. Similarly, $\Delta_z$ and $\Delta_\theta$ were defined as $24\lambda v$ and $4°v$. For instance, if we take $v = 0.5$, this means each diffractive layer can independently/randomly shift in both $x$ and $y$ axes within a range of $(-0.335\lambda, 0.335\lambda)$. In addition, their location over the $z$ direction and their in-plane orientation can randomly change within the ranges of $(-12\lambda, 12\lambda)$ and $(-2°, 2°)$, respectively (see the Methods section for more details).

To better highlight the impact of these misalignments and demonstrate the efficacy of the v-D²NN framework, we trained a new nonvaccinated, i.e., $v_{tr} = 0$, diffractive permutation network that can all-optically realize a given permutation matrix, $\boldsymbol{P}$, representing 10K intensity interconnections between 100 input and 100 output pixels of size $4\lambda \times 4\lambda$. The error-free training model of this diffractive network with $v_{tr} = 0$ implicitly assumes that when the resulting diffractive network is fabricated, the system conditions will exactly match the ideal settings regarding the 3D locations of the layers and their in-plane orientations. With an architecture identical to the one shown in Fig. 1, containing $N = 200K \gg N_i N_o$ diffractive neurons, this diffractive network can all-optically approximate the permutation matrix, $\boldsymbol{P}$, with an MSE of $1.45 \times 10^{-6}$ in the absence of any misalignment errors, i.e., $v_{test} = 0$ (see the green curve in Fig. 4B). However, when there is some discrepancy between the training and testing conditions, i.e., $v_{test} > 0$, the optically implemented forward transformation, $\boldsymbol{P}_{D^2NN}$, starts to deviate from the desired operation $\boldsymbol{P}$. For instance, at $v_{test} = 0.125$, the transformation error, $\left\|\boldsymbol{P}_{D^2NN} - \boldsymbol{P}\right\|$, can be computed as $3.1 \times 10^{-3}$. This negative impact of the physical misalignments on the performance of a *nonvaccinated* diffractive network can also be seen in Fig. 4A (green curve), which demonstrates the SSIM values achieved by this diffractive network for recovering permuted EMNIST images under different levels of misalignments. The high-quality of the image recovery (see Fig. 4C) at $v_{test} = 0$ quantified with an SSIM of 0.99 deteriorates under the presence of misalignments, highlighted by the SSIM value falling to 0.49 and 0.30 at $v_{test} = 0.125$ and $v_{test} = 0.25$, respectively.

Unlike the nonvaccinated design, the vaccinated diffractive permutation networks can maintain their approximation capacity and accuracy under erroneous testing conditions as shown in Figs. 4A-B. For instance, the SSIM value of 0.49 attained by the nonvaccinated diffractive network for the misalignment uncertainty set by $v_{test} = 0.125$, increases to 0.88 in the case of a diffractive permutation network trained with $v_{tr} = 0.25$ (red curve in Fig. 4A). The difference between the image recovery performances of the vaccinated and the nonvaccinated diffractive network designs increases further as the misalignment levels increase during the blind testing. While the nonvaccinated diffractive network can only achieve SSIM values of 0.3 and 0.19 at $v_{test} = 0.25$ and $v_{test} = 0.375$, respectively, the output images synthesized by the vaccinated design ($v_{tr} = 0.25$) reveals SSIM values of 0.8 at $v_{test} = 0.25$ and 0.64 at $v_{test} = 0.375$ (see Fig. 4D). A similar conclusion can also be drawn from Fig. 4B, demonstrating the MSE values between the desired permutation matrix, $\boldsymbol{P}$, and its optically realized counterpart, $\boldsymbol{P}_{D^2NN}$. The transformation errors, $\left\|\boldsymbol{P}_{D^2NN} - \boldsymbol{P}\right\|$, of the vaccinated diffractive network ($v_{tr} = 0.25$) at $v_{test} = 0.125$ and at $v_{test} = 0.25$ were computed as $5.15 \times 10^{-4}$ and $1.2 \times 10^{-3}$, respectively, which are 5-10 times smaller compared to the MSE values provided by the nonvaccinated diffractive design at the same misalignment levels. Supplementary Fig. S2 further illustrates the error maps between $\boldsymbol{P}$ and $\boldsymbol{P}_{D^2NN}$ realized by the nonvaccinated and vaccinated diffractive permutation networks at different misalignment levels.

The compromise for this misalignment robustness comes in the form of a reduction in the peak performance. While the nonvaccinated diffractive network can solely focus on realizing the given permutation operation with the highest quality and approximation accuracy, the vaccinated diffractive network designs partially allocate their degrees-of-freedom to building up resilience against physical misalignments. For example, while the peak SSIM achieved by the nonvaccinated diffractive network is 0.99, it is 0.88 for the diffractive permutation network vaccinated with $v_{tr} = 0.25$. The key difference, on the other hand, is that the better performance of the nonvaccinated diffractive network is sensitive to the physical implementation errors, while the vaccinated diffractive permutation networks can realize the desired input-output interconnects over a larger range of fabrication errors or tolerances. A comparison between the diffractive layer patterns of the nonvaccinated and vaccinated diffractive permutation networks shown in Figs. 4C and 4D, respectively, also reveals that the vaccination strategy results in smoother light modulation patterns; in other words, the material thickness values over the neighboring diffractive neurons partially lose their independence and become correlated, causing a reduction in the number of independent degrees-of-freedom in the system.

## 2.3 Experimental demonstration of a diffractive permutation network

To experimentally demonstrate the success of the presented diffractive permutation interconnects, we designed a 3-layer diffractive permutation network achieving the desired (randomly generated) intensity shuffling operation with $N_i = N_o = 5x5$, optically synthesizing 625 connections between the input and output FOVs; this network was designed to operate at 0.4 THz, corresponding to ~0.75 mm in wavelength. During the training, the forward model of this diffractive permutation network was vaccinated with $v_{tr} = 0.5$ against the 4 error sources as detailed in Section 2.2 including the 3D location of each diffractive layer and the in-plane rotation angle around the optical axis. In addition to these misalignment components, we also vaccinated this diffractive network model against unwanted material thickness variations that could arise due to the limited lateral and axial resolution of our 3D printer (see the Methods section for details). To compensate for the reduction in the degrees-of-freedom due to the vaccination scheme, the number of phase-only diffractive features in the permutation network was selected to be $N_L = 10$K diffractive neurons per layer. Therefore, each diffractive layer shown in Fig. 5A contains 100×100 phase-only diffractive neurons of size ~$0.67\lambda \times 0.67\lambda$. Compared to the diffractive surfaces shown in Figs. 1-4, the layers of our experimental system were set to be 2-times smaller in both the $x$ and $y$ directions to keep the layer-to-layer distances smaller while maintaining the level of optical connectivity between the successive diffractive surfaces (see Fig. 5B). Figures 5C and 5D illustrate the 3D printed diffractive permutation network and the schematic of our experimental setup (see the Methods section for details).

Figure 6A illustrates the targeted 25×25 permutation matrix ($P$) that is randomly generated and the numerically predicted $P_{D^2NN}$ along with the absolute difference map between these two matrices. According to the numerical forward model of the trained diffractive network shown in Fig. 5, the transformation error between the $P$ and $P_{D^2NN}$, i.e., $\|P_{D^2NN} - P\|$ is equal to $5.99 \times 10^{-4}$ under error-free conditions, i.e., $v_{test} = 0$. Furthermore, the forward model of the trained diffractive permutation network shown in Fig. 5 provides 17.87 dB PSNR on average for the test letters 'U', 'C', 'L' and 'A', as depicted in Fig. 6B. A visual comparison between the numerically predicted and the experimentally measured output images of these 4 input letters (which were never seen by the network before) demonstrates the accuracy of the forward training and testing models as well as the success of the presented diffractive permutation network design. Interestingly, the PSNR of the experimentally measured images was observed to be higher, 19.54 dB, compared to the numerically predicted value, 17.87 dB. Our numerical study reported in Fig. 4 suggests that this can be explained based on the vaccination range used during the training and the amount physical error in the system testing. For instance, the SSIM value achieved by the vaccinated diffractive network trained with $v_{tr} = 0.5$ (yellow curve) at relatively lower physical misalignment levels, e.g., $v_{test} = 0.125$, is higher compared to its performance under the ideal conditions, i.e., $v_{test} = 0.0$, as depicted in Fig. 4A.

# 3 Discussion

Beyond optomechanical error sources and fabrication tolerances, another factor that might potentially hinder the utilization of diffractive permutation networks in practical applications is the output diffraction efficiency. For instance, the diffraction efficiency of the 5-layer network shown in Fig. 1 is ~0.004% which might be very low for some applications. On the other hand, this can be significantly increased by using an additional loss term, penalizing the poor diffraction efficiency of the network (see the Methods section for details). Supplementary Fig. S3 demonstrates 5-layer diffractive networks that are designed to optically realize 0.16 million interconnections between the input and output FOVs with increased diffraction efficiencies, and compares their performance in terms of SSIM values achieved. The training of these diffractive network models is based on a loss function in the form of a linear combination of two different penalty terms, $\mathcal{L}' = \mathcal{L} + \gamma \mathcal{L}_e$, where $\mathcal{L}$ is a structural loss term enforcing transformation quality/accuracy and $\mathcal{L}_e$ is the diffraction efficiency related penalty term promoting efficient solutions (see the Methods section). As a general trend, the diffraction efficiency of the underlying diffractive network model increases as a function of the weight ($\gamma$) of the efficiency penalty term in the loss function. However, since the number of diffractive neurons, hence, the degrees-of-freedom in these diffractive network models is very close to $N_iN_o$, the diffraction efficiency either does not improve beyond a certain value or the evolution of the diffractive layers starts to solely focus on the efficiency instead of the desired permutation operation resulting in low performance designs. This unstable behavior can be observed specifically when $0.235 < \gamma < 0.24$. On the other hand, as in the case of vaccinated diffractive network models, if the diffractive network architecture contains $N \gg N_i N_o$ diffractive neurons, then this instability vanishes, providing significant improvements in the output diffraction efficiency without sacrificing the performance of the all-optical permutation operation. For instance, the 3D printed diffractive permutation network depicted in Fig. 6 was trained based on $\mathcal{L}'$ with $\gamma = 0.15$ and it provides 2.45% output diffraction efficiency, despite the fact that 89.37% of the incident power at the input plane is lost due to the absorption of the 3D printing material. With weakly absorbing transparent materials used as part of the diffractive network fabrication, a significantly larger output efficiency can be achieved.

In summary, we showed that the diffractive networks can optically implement intensity permutation operations between their input and output apertures based on phase-only light modulation surfaces with $N \geq N_i N_o$ diffractive neurons. Due to the nonlinear nature of the intensity operation, it is crucial to use training input intensity patterns with different levels of sparsity to prevent any type of data-specific overfitting during the training phase. Diffractive permutation networks with $N > N_i N_o$ demonstrate increased generalization capability, synthesizing more accurate outputs with $\left\| \boldsymbol{P_{D^2NN}} - \boldsymbol{P} \right\| \approx 0$. By using $N > N_i N_o$ one can also design misalignment and fabrication error insensitive, power-efficient diffractive permutation networks, which could play a major role in practical applications, e.g., 6G wireless networks, computational cameras, etc. Finally, the incorporation of dynamic spatial light modulators to replace some of the diffractive layers in a given design can be used to reconfigure, on demand, the all-optically performed diffractive transformation.

# 4 Methods

## 4.1 Experimental setup

According to the schematic diagram of our experimental setup shown in Fig. 5D, the THz wave incident on the input FOV of the diffractive network was generated using a horn antenna attached to the source WR2.2 modulator amplifier/multiplier chain (AMC) from Virginia Diode Inc. (VDI). A 10 dBm RF input signal at 11.111 GHz ($f_{RF1}$) at the input of the AMC was multiplied 36 times to generate a continuous-wave (CW) radiation at 0.4 THz, corresponding to ~0.75 mm in wavelength. The output of the AMC was modulated with 1 kHz square wave to resolve low-noise output data through lock-in detection. Since we did not use any collimating optics in our setup, the distance between the input plane of the 3D-printed diffractive optical network and the exit aperture of the horn antenna was set to be ~60 cm approximating a uniform plane wave over the $40\lambda \times 40\lambda$ input FOV. At the output plane of the diffractive optical network, the diffracted THz light was collected using a single-pixel Mixer/AMC from Virginia Diode Inc. (VDI). During the measurements, the detector received a 10 dBm sinusoidal signal at 11.083 GHz serving as a local oscillator for mixing, and the down-converted signal was at 1GHz. The $40\lambda \times 40\lambda$ output FOV was scanned by placing the single-pixel detector on an XY stage that was built by combining two linear motorized stages (Thorlabs NRT100). At each scan location, the down-converted signal coming from the single-pixel detector was fed to low-noise amplifiers (Mini-Circuits ZRL-1150-LN+) with a gain of 80 dBm and a 1 GHz (+/-10 MHz) bandpass filter (KL Electronics 3C40-1000/T10-O/O) that erases the noise components coming from unwanted frequency bands. Following the amplification and filtering, the measured signal passed through a tunable attenuator (HP 8495B) and a low-noise power detector (Mini-Circuits ZX47-60). Finally, the output voltage value was generated by a lock-in amplifier (Stanford Research SR830). The modulation signal was used as the reference signal for the lock-in amplifier and accordingly, we performed a calibration to convert the lock-in amplifier readings at each scan location to linear scale. During our experiments, the scanning step size at the output plane was set to be ~$\lambda$ in x and y directions. The smallest pixel of the experimentally targeted permutation grid, i.e., the desired resolution of the diffractive permutation operation was taken as $8\lambda \times 8\lambda$ during the training, corresponding to $5 \times 5$ discrete input and output signals. Therefore, the output signal measured for each input object was integrated over a region of $8\lambda \times 8\lambda$ per pixel, resulting in the measured images shown in Fig. 6B.

A 3D printer, Objet30 Pro, from Stratasys Ltd., was used to fabricate the layers of the diffractive permutation network shown in Fig. 5C as well as the layer holders. The active modulation area of our 3D printed diffractive layers was 5 cm × 5 cm (~66.66$\lambda$ × ~66.66$\lambda$) containing $100 \times 100$, i.e., 10K, diffractive neurons. These modulation surfaces were printed as insets surrounded by a uniform slab of printing material with a thickness of 2.5 mm and the total size of each printed layer including these uniform regions was 6.2 cm × 6.2 cm. Following the 3D printing, these additional surrounding regions were coated with aluminum to block the propagation of the light over these areas minimizing the contamination of the output signal with unwanted scattered light.

## 4.2 Training forward model of diffractive permutation networks

### 4.2.1 Optical forward model

The material thickness, $h$, was selected as the physical parameter controlling the complex-valued transmittance values of the diffractive layers of our design. Based on the complex-valued refractive index of the diffractive material, $\tau = n + j\kappa$, the corresponding transmission coefficient of a diffractive neuron located on the $l^{th}$ layer at a coordinate of $(x_q, y_q, z_l)$ is defined as,

$$t(x_q, y_q, z_l) = \exp\left(\frac{-2\pi\kappa h(x_q, y_q, z_l)}{\lambda}\right) \exp\left(\frac{-j2\pi(n - n_m)h(x_q, y_q, z_l)}{\lambda}\right) \tag{1}$$

where $n_m = 1$ denotes the refractive index of the propagation medium (air) between the layers. The real and imaginary parts of the 3D printing material were measured experimentally using a THz spectroscopy system and they were revealed as $n = 1.7227$ and $\kappa = 0.031$ at 0.4 THz.

The optical forward model of the presented diffractive networks relies on the Rayleigh-Sommerfeld theory of scalar diffraction to represent the propagation of light waves between the successive layers. According to this diffraction formulation, the free-space can be interpreted as a linear, shift-invariant operator with the impulse response,

$$w(x, y, z) = \frac{z}{r^2}\left(\frac{1}{2\pi r} + \frac{n}{j\lambda}\right)\exp\left(\frac{j2\pi nr}{\lambda}\right) \tag{2}$$

where $r = \sqrt{x^2 + y^2 + z^2}$. Based on Eq. 2, $q^{th}$ diffractive neuron on the $l^{th}$ layer, at $(x_q, y_q, z_l)$, can be interpreted as the source of a secondary wave generating the field at $(x, y, z)$ in the form of,

$$w_q^l(x, y, z) = \frac{z - z_l}{\left(r_q^l\right)^2}\left(\frac{1}{2\pi r_q^l} + \frac{n}{j\lambda}\right)\exp\left(\frac{j2\pi nr_q^l}{\lambda}\right). \tag{3}$$

The parameter $r_q^l$ in Eq. 3 is expressed as $\sqrt{\left(x - x_q\right)^2 + \left(y - y_q\right)^2 + (z - z_l)^2}$. When each diffractive neuron on layer $l$ generates the field described by Eq. 3, the light field incident on the $p^{th}$ diffractive neuron on the $(l + 1)^{th}$ layer at $(x_p, y_p, z_{l+1})$ is the linear superposition of the all the secondary waves generated by the previous layer $l$, i.e., $\sum_q A_q^l w_q^l(x_p, y_p, z_{l+1})$, where $A_q^l$ is the complex amplitude of the wave field right after the $q^{th}$ neuron of the $l^{th}$ layer. This field is modulated by the multiplicative complex-valued transmittance of the diffractive unit at $(x_p, y_p, z_{l+1})$, creating the modulated field $t(x_p, y_p, z_{l+1})\sum_q A_q^l w_q^l(x_p, y_p, z_{l+1})$. Based on this new modulated field, a new secondary wave,

$$u_p^{l+1}(x, y, z) = w_p^{l+1}(x, y, z)t(x_p, y_p, z_{l+1})\sum_q A_q^l w_q^l(x_p, y_p, z_{l+1}), \tag{4}$$

is generated. The outlined successive modulation and secondary wave generation processes occur until the waves propagating through the diffractive network reach to the output plane. Although, the forward optical model described by Eqs. 1-4 is given over a continuous 3D coordinate system, during our deep learning-based training of the presented diffractive permutation networks, all the wave fields and the modulation surfaces were represented based on their discrete counterparts with a spatial sampling rate of ~$0.67\lambda$ on both x and y axes, that is also equal to the size of a diffractive neuron.

### 4.2.2 Physical architecture of the diffractive permutation networks and training loss functions

The size of the output and input FOVs of the presented diffractive permutation networks were both set to be $40\lambda \times 40\lambda$, defining a unit magnification optical permutation operation. Note that the unit magnification is not a necessary condition for the success of the forward operation of diffractive optical interconnects but rather a design choice. Without loss of generality, the output FOV can be defined centered around the origin, $(0,0)$, i.e., $-20\lambda < x, y < 20\lambda$. The dimensions of the diffractive layers was taken as 133.3λ×133.3λ for the diffractive permutation networks presented in Figs. 1-4 and Supplementary Fig. S3, and in all these diffractive network architectures the layer-to-layer distances were taken as 120λ. The axial distance between the 1st diffractive layer and the input FOV was set to be 53.3λ that is also equal to the axial distance from the last diffractive layer to the output plane, preserving the symmetry of the system on the longitudinal axis. In the case of our experimentally validated diffractive design (Fig. 5), on the other hand, the active modulation surface of the fabricated diffractive layers extends 66.7λ on both $x$ and $y$ directions. Accordingly, the layer-to-layer distances were taken as 60λ while the remaining distances were kept equal to 53.3λ.

During the deep learning-based training of all of these diffractive permutation networks, the wave fields and the propagation functions depicted in Eqs. 2-4 were sampled at a rate of ~0.67λ that is also equal to the size of the smallest diffractive units on the modulation surfaces constituting the presented diffractive networks. At this spatial sampling rate, the input and output intensity patterns were represented as 2D discrete vectors of size 60×60 denoted by $I_{in}[m,n]$ and $I_{out}[m,n]$, respectively, with $m = 1,2,3,...,60$ and $n = 1,2,3,...,60$. The underlying complex-valued wave fields can be written as $U_{in}[m,n] = \sqrt{I_{in}[m,n]}e^{j\emptyset_{in}[m,n]}$ and $U_{out}[m,n] = \sqrt{I_{out}[m,n]}e^{j\emptyset_{out}[m,n]}$. In our forward model, we assumed that the input light has constant phase front, i.e., $\emptyset_{in}[m,n]$ is taken as an arbitrary constant within the input field-of-view. In alternative implementations, without loss of generality, the diffractive permutation network can be trained with any arbitrary function of $\emptyset_{in}[m,n]$, achieving the same output accuracy levels $\left\| \boldsymbol{P_{D^2NN}} - \boldsymbol{P} \right\| \approx 0$ using $N \geq N_i N_o$.

While the light fields, the diffractive layers and the impulse response of the free-space were all sampled at a rate of ~0.67λ, the spatial grid/pixel size of a given desired permutation operation was set to be larger. Specifically, the permutation pixel size was taken as 2λ×2λ for the diffractive networks shown in Figs. 1-3. On the other hand, the input and output pixel size was chosen as 4λ×4λ for the vaccinated and nonvaccinated diffractive permutation networks shown in Fig. 4; and finally, the pixel size was set to be 8λ×8λ for the fabricated diffractive permutation network model depicted in Fig. 5.

To train the presented diffractive permutation networks, a structural loss function, $\mathcal{L}$, in the form of MSE was used.

$$\mathcal{L} = \frac{1}{S}\sum_{s=1}^{S}|\boldsymbol{P}I_{in}[s] - \sigma I_{out}[s]|^2,$$ (5)

In Eq. 5, $I_{in}[s]$ and $I_{out}[s]$ denote the lexicographically ordered vectorized counterparts of the input intensity pattern, i.e., $\text{vec}(I_{in}[q,p])$, and the output intensity pattern, i.e., $vec(I_{out}[q,p])$, and $\boldsymbol{P}$ represents the desired permutation matrix to be performed all-optically. As depicted in Eq. 5, the output intensity pattern $I_{out}[s]$ or $I_{out}[q,p]$ was scaled by a constant $\sigma$ that was calculated at each training iteration as,

$$\sigma = \frac{\frac{1}{S}\sum_{s=1}^{S}I_{in}[s]I_{out}[s]}{\frac{1}{S}\sum_{s=1}^{S}I_{out}[s]^2}.$$ (6)

To improve the diffraction efficiency of diffractive permutation networks, we defined another loss function, $\mathcal{L}'$, that is a linear combination of two penalty terms, $\mathcal{L}' = \mathcal{L} + \gamma \mathcal{L}_e$, where $\mathcal{L}$ corresponds to the structural loss defined in Eq. 5. $\mathcal{L}_e$ is the penalty term that promotes higher diffraction efficiency at the output of diffractive networks, and it was defined as, $\mathcal{L}_e = e^{-\eta}$, where,

$$\eta = \frac{\sum_{s=1}^{S}I_{out}[s]}{\sum_{s=1}^{S}I_{in}[s]} \times 100.$$ (7)

The diffractive permutation networks presented in Figs. 1-4 were trained based on $\mathcal{L}'$ with $\gamma = 0$; however, the experimentally demonstrated diffractive permutation network model was trained with $\gamma = 0.15$, resulting in an output diffraction efficiency of 2.45% (which includes a material absorption loss of 89.37%). Supplementary Figure S3 further demonstrates the diffraction efficiency and the SSIM values provided by various diffractive permutation network models trained with different $\gamma$ values.

The supervised deep learning-based training of the presented diffractive permutation networks evaluates the loss function $\mathcal{L}'$ for a batch of randomly generated input patterns, computes the mean gradient and updates the learnable, auxiliary variables, $h_a$, that determine the material thickness over each diffractive neuron, $h$, through the following relation,

$$h(h_a) = \frac{\sin(h_a) + 1}{2}(h_m - h_b) + h_b$$ (8)

where $h_m$ and $h_b$ denote the maximum modulation thickness and the base material thickness, respectively. For all the diffractive permutation networks presented in Figs. 1-4 and Supplementary Fig. S3, $h_m$ was taken as 2λ. In the design of the 3D-printed diffractive permutation network, however, $h_m$ was set to be 1.66λ to restrict the material thickness contrast between the neighboring diffractive features. The value of $h_b$ was taken as 0.66$\lambda$ for all the presented designs including the fabricated diffractive network.

### 4.2.3 Computation of $P_{D^2NN}$, optical transformation errors and performance quality metrics

For a given diffractive permutation network design trained to optically implement a permutation matrix $P$ of size $N_i \times N_o$, there are two different ways to compute the permutation operation predicted by its numerical forward model. The first way is to propagate $N$ different randomly generated independent inputs with $N \geq N_i N_o$ and solve a linear system of equations for revealing the entries of $P_{D^2NN}$. Alternatively, each input pixel at the input FOV can be turned on sequentially and the output intensity pattern synthesized by the diffractive optical permutation network as a response to each pixel provides one unique column of $P_{D^2NN}$. These two procedures, in general, result in two different $P_{D^2NN}$ matrices that closely resemble each other. We opted to use the latter procedure due to its simplicity, which turn on each input pixel one at a time and records the corresponding output intensity pattern, which, after vectorization, represents a column of $P_{D^2NN}$. Following the calculation of $P_{D^2NN}$ predicted by the forward model of a trained diffractive permutation network, it was scaled with a multiplicative constant, $\sigma_P$, to account for the optical losses:

$$\sigma_P = \frac{\frac{1}{N_i N_o} \sum_{n_i}^{N_i} \sum_{n_o}^{N_o} P_{D^2NN}[n_i, n_o] P[n_i, n_o]}{\frac{1}{N_i N_o} \sum_{n_i}^{N_i} \sum_{n_o}^{N_o} P_{D^2NN}[n_i, n_o]^2}. \tag{9}$$

The all-optical transformation error, $\left\| P - P_{D^2NN} \right\|^2$ can be computed based on,

$$\left\| P - P_{D^2NN} \right\|^2 = \frac{1}{N_i N_o} \sum_{n_i}^{N_i} \sum_{n_o}^{N_o} |\sigma_P P_{D^2NN}[n_i, n_o] - P[n_i, n_o]|^2. \tag{10}$$

Denoting the lexicographically ordered vectorized version of a 2D input intensity pattern with $I_{in}[s]$, the ground truth output intensity can be found by $P I_{in}[s]$. The PSNR between this ground-truth vector and the output vector synthesized by the forward optical operation of a given, trained diffractive network, $I_{out}[s]$, can be calculated as,

$$PSNR = 20 log_{10} \left( \frac{1}{\sqrt{\sum_s |P I_{in}[s] - \sigma I_{out}[s]|^2}} \right), \tag{11}$$

where $\sigma$ is the multiplicative constant defined in Eq. 6. The SSIM values were calculated based on the built-in function in TensorFlow, i.e., tf.image.ssim, where the two inputs were 2D versions of $P I_{in}[s]$ and $I_{out}[s]$, representing the ground-truth image and the permuted, all-optical output signal, respectively. All the parameters of tf.image.ssim were taken equal to default values, except that the size of the Gaussian filter was set to be 5×5, instead of 11×11, and the width of the Gaussian filter was set to be 0.75.

### 4.2.4 Vaccination of diffractive permutation networks

v-D$^2$NN framework aims to design diffractive optical networks that are resilient against physical error sources, e.g., misalignments, by modeling these factors as random variables and incorporating them into the forward training model. In the training forward model of the vaccinated diffractive networks shown in Fig. 4, 4 physical error components were modeled representing the misalignment of each diffractive layer with respect to their ideal location and orientation/angle. The first 3 components represent the statistical variations in the location of each diffractive layer in 3D space. Let the ideal location of a diffractive layer, $l$, be denoted by the vector $X^l = (x_l, y_l, z_l)$, then at each training iteration $i$, v-D$^2$NN framework perturbs $X^l$ with a random displacement vector, $D^{l,i} = \left( D_x{}^{l,i}, D_y{}^{l,i}, D_z{}^{l,i} \right)$. The components of this 3D displacement vector were defined as uniformly distributed, independent random variables, i.e.,

$$D_x{}^{l,i} \sim U(-\Delta_x, \Delta_x) \tag{12a}$$

$$D_y{}^{l,i} \sim U(-\Delta_y, \Delta_y) \tag{12b}$$

$$D_z{}^{l,i} \sim U(-\Delta_y, \Delta_y) \tag{12c}$$

During the training, for each batch of input images, the 3D displacement vector $D^{l,i}$ is updated and accordingly, the location of the layer $l$ is set to be $X^{l,i} = X^l + D^{l,i}$, building up robustness to physical misalignments.

Beyond the displacement of diffractive layers, the physical forward model of a diffractive network is also susceptible to variations in the orientation of the diffractive layers. Ideally, one should include all 3 rotational components, yaw, pitch and roll, however, in this study we only considered the yaw component since in our experimental systems, the pitch and the roll can be controlled with a high precision. The random angle representing the rotation of a diffractive layer $l$ around the optical axis was defined as $D_\theta^{l,i} \sim U(-\Delta_\theta, \Delta_\theta)$. With 3 shift components depicted in Eq. 12 and the statistical yaw variation modeled through $D_\theta^{l,i}$, the vaccinated diffractive networks shown in Fig. 4 were trained to build resilience against these 4 misalignment components. The values of $\Delta_x$, $\Delta_y$, $\Delta_z$ and $\Delta_\theta$ determining the misalignment tolerance range were defined as a function a common variable $v$, i.e., $\Delta_x = \Delta_y = 0.67\lambda v$, $\Delta_z = 24\lambda v$ and $\Delta_\theta = 4°$.

For the design of the experimentally validated diffractive permutation network, on top of these 4 optomechanical error components (with $v = 0.5$), we also modeled fabrication errors in the form of statistical variations of the material thickness over each diffractive neuron ($h$). Hence, at a given iteration, $i$, the material thickness values over each diffractive unit $h(h_a)$, defined in Eq. 8 was perturbed through $h^i(h_a) = h(h_a) + D_h^i$, where $D_h^i \sim U(-0.025h_m, 0.025h_m)$. Stated differently, the fabricated diffractive layers shown in Fig. 5 were designed to be resilient against physical errors on the material thickness values over the diffractive neurons within a range $[-0.0415\lambda, 0.0415\lambda]$.

### 4.2.5 Training details

The deep learning-based training of the diffractive permutation networks were implemented using Python (v3.6.5) and TensorFlow (v1.15.0, Google Inc.). The backpropagation updates were calculated using the Adam optimizer[79], and its parameters were taken as the default values in TensorFlow and kept identical in each model. The learning rates of the diffractive optical networks were set to be 0.001. The training batch size was taken as 75 during the deep learning-based training of the presented diffractive permutation networks. For the training of the diffractive permutation networks, we generated ~4.7 million random intensity patterns, providing us 93750 iterations/error-backpropagation updates per epoch. The training of a 5-layer diffractive permutation network with 40K diffractive neurons per layer for 5 epochs using this randomly created training dataset takes approximately 4 days using a computer with a GeForce GTX 1080 Ti Graphical Processing Unit (GPU, Nvidia Inc.) and Intel® Core ™ i7-8700 Central Processing Unit (CPU, Intel Inc.) with 64 GB of RAM, running Windows 10 operating system (Microsoft).

# Figures and Figure Captions

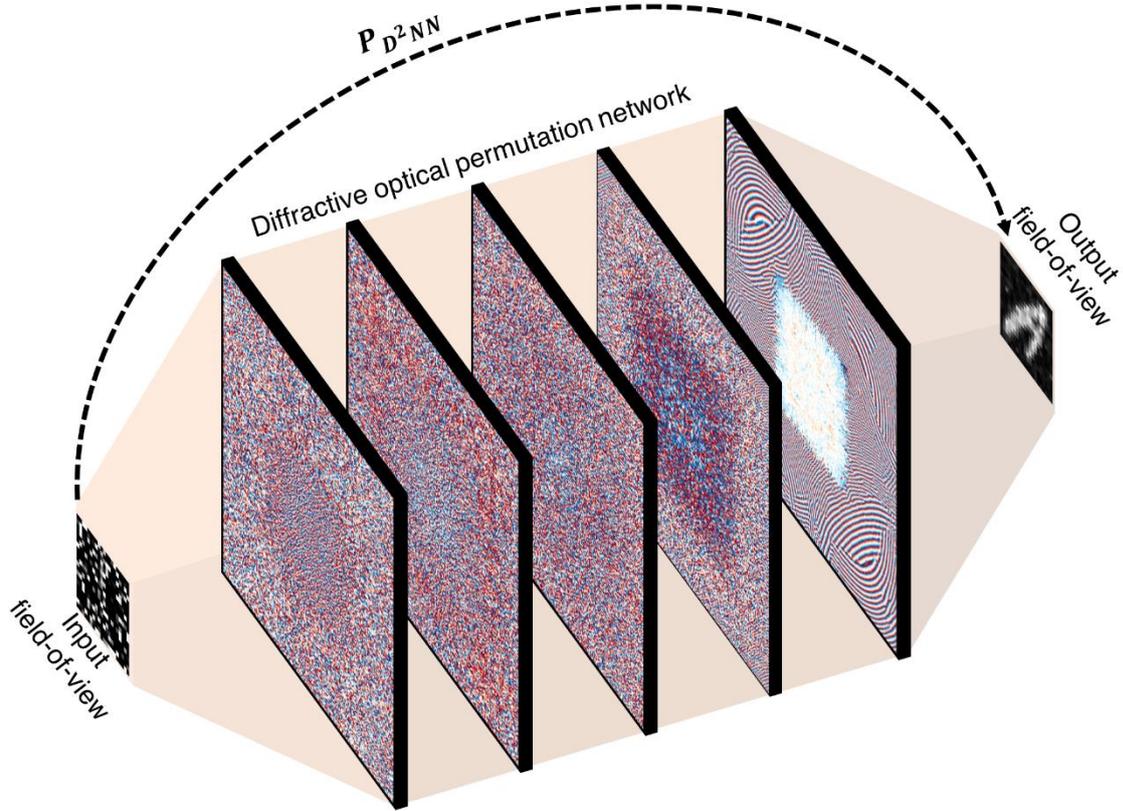

**Fig. 1: The schematic of a 5-layer diffractive permutation network, all-optically realizing 0.16 million interconnects between an input and output field-of-view.** The presented diffractive permutation network was trained to optically realize an arbitrarily-selected permutation operation between the light intensities over $N_i = 400 = 20 \times 20$ input and $N_o = 400 = 20 \times 20$ output pixels, establishing $N_i N_o = 0.16$ million desired interconnections based on 5 phase-only diffractive layers, each containing 40K ($200 \times 200$) diffractive neurons/features.

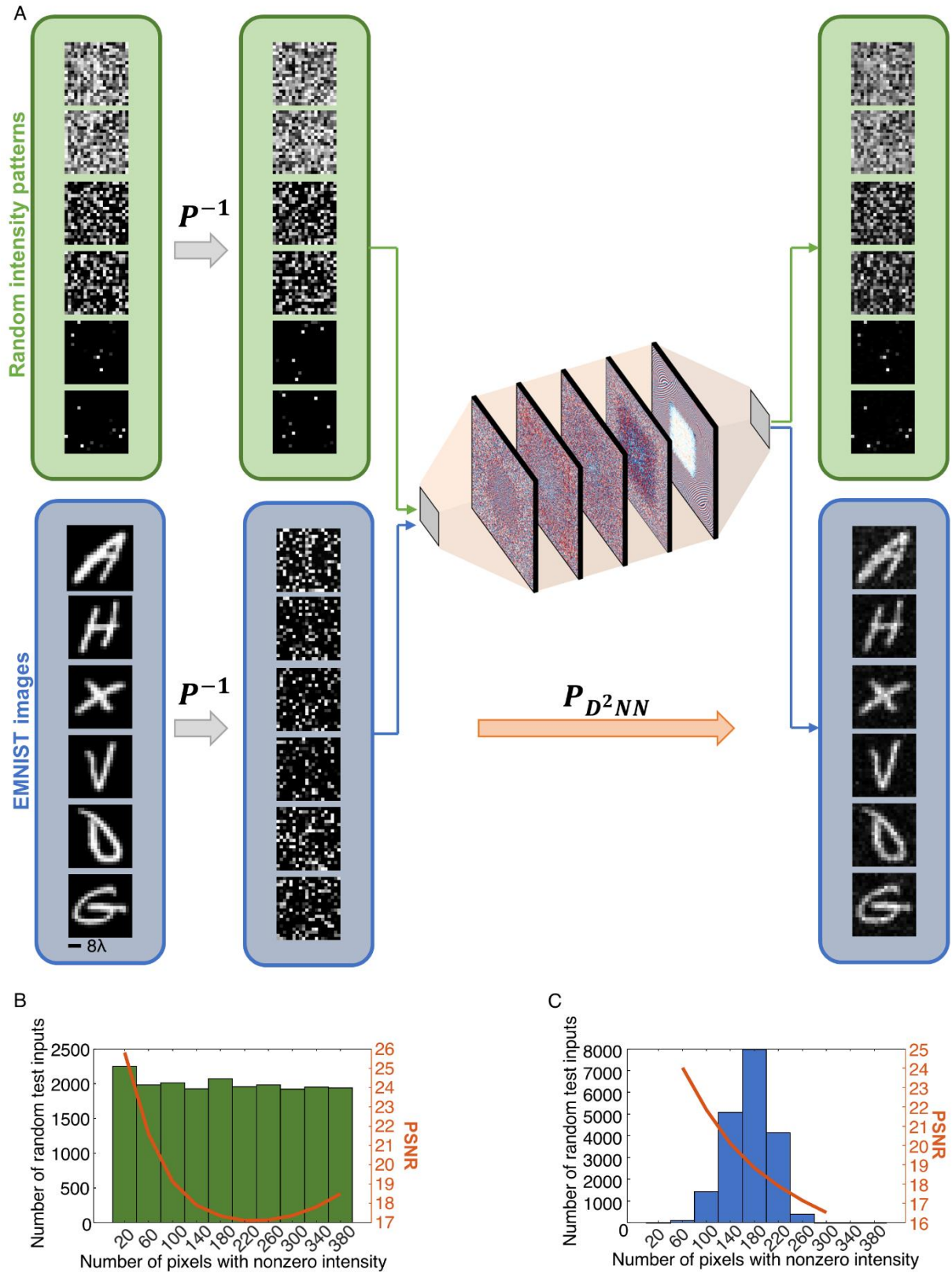

**Fig. 2: Input-output intensity pairs generated by the diffractive permutation network shown in Fig. 1.** A The diffractive permutation network shown in Fig. 1 was tested on two different datasets. The first blind testing dataset contains 20K randomly generated inputs. 6 examples from this randomly created testing data are shown here for demonstrating input-output intensity pairs with low, moderate and high signal sparsity levels. Beyond successfully permuting randomly generated intensity patterns, the performance of the diffractive permutation network was also quantified using permuted EMNIST images. None of these test samples were used in the training phase. B Output intensity image PSNR with respect to the ground truth intensity patterns as a function of the input signal sparsity in randomly generated test dataset. C Same as B, except for EMNIST test images.

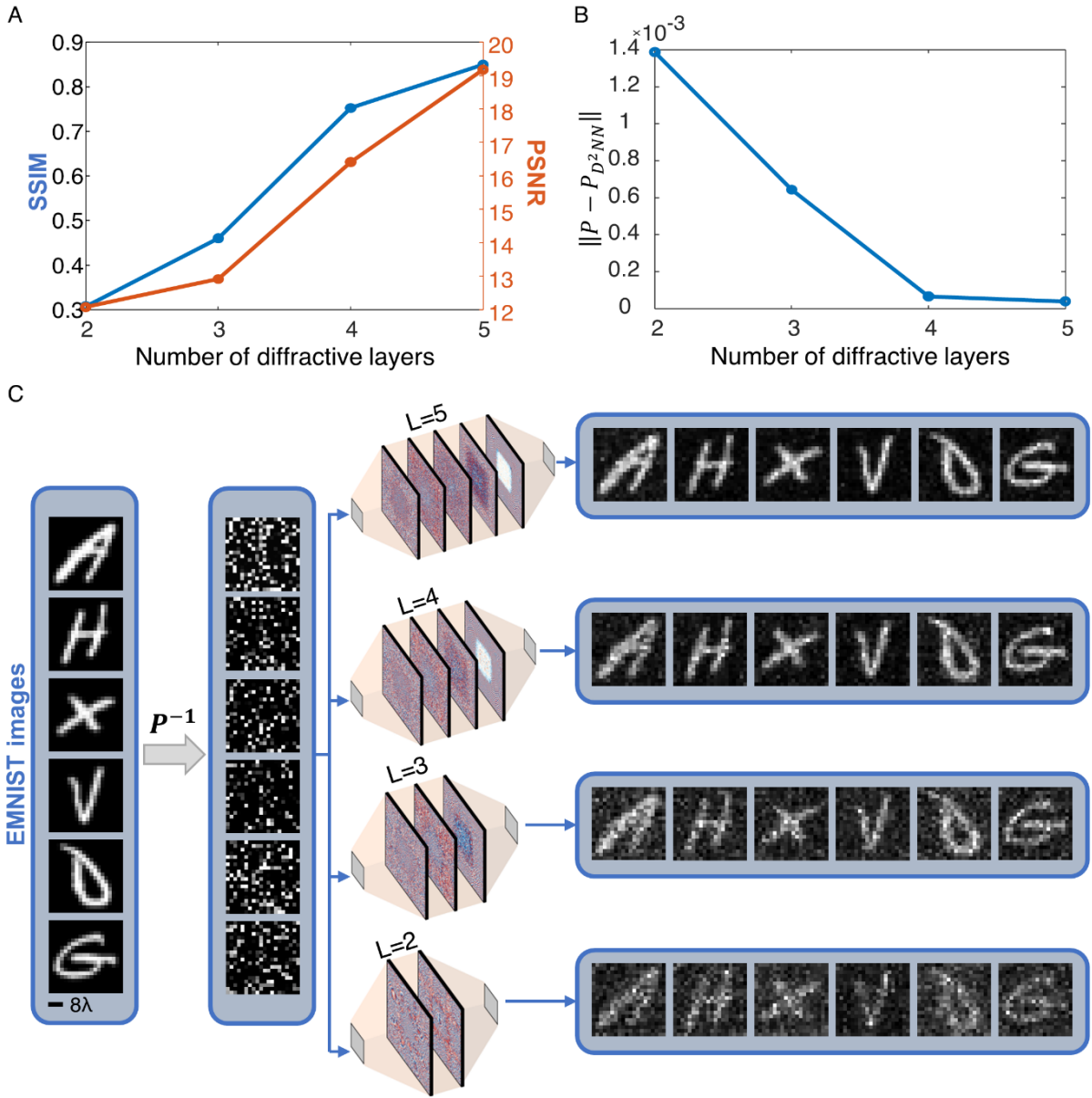

**Fig. 3: The impact of the number of diffractive layers on the approximation accuracy of D²NN for a given intensity permutation operation.** A The average SSIM and PSNR values achieved by the diffractive permutation network designs based on L=2, L=3, L=4 and L=5 diffractive layers, containing 200×200, i.e., 40K, phase-only diffractive neurons/features per layer for the task of optically recovering permuted EMNIST images. B The transformation error between the desired intensity permutation ($P$) and its optically realized counterpart ($P_{D^2NN}$) for the diffractive networks with L=2, L=3, L=4 and L=5 diffractive layers. The transformation error decreases as a function of the number of layers in the diffractive network architecture. The L=4-layer diffractive permutation network design represents a critical point as it matches the space-bandwidth product requirement of the desired permutation operation, i.e., $N = N_i N_o = 4 \times 40K = 160K$, and further increasing the number of layers to L=5 brings only a minor improvement. C Examples of EMNIST test images demonstrating the performance of the diffractive permutation networks as a function of L.

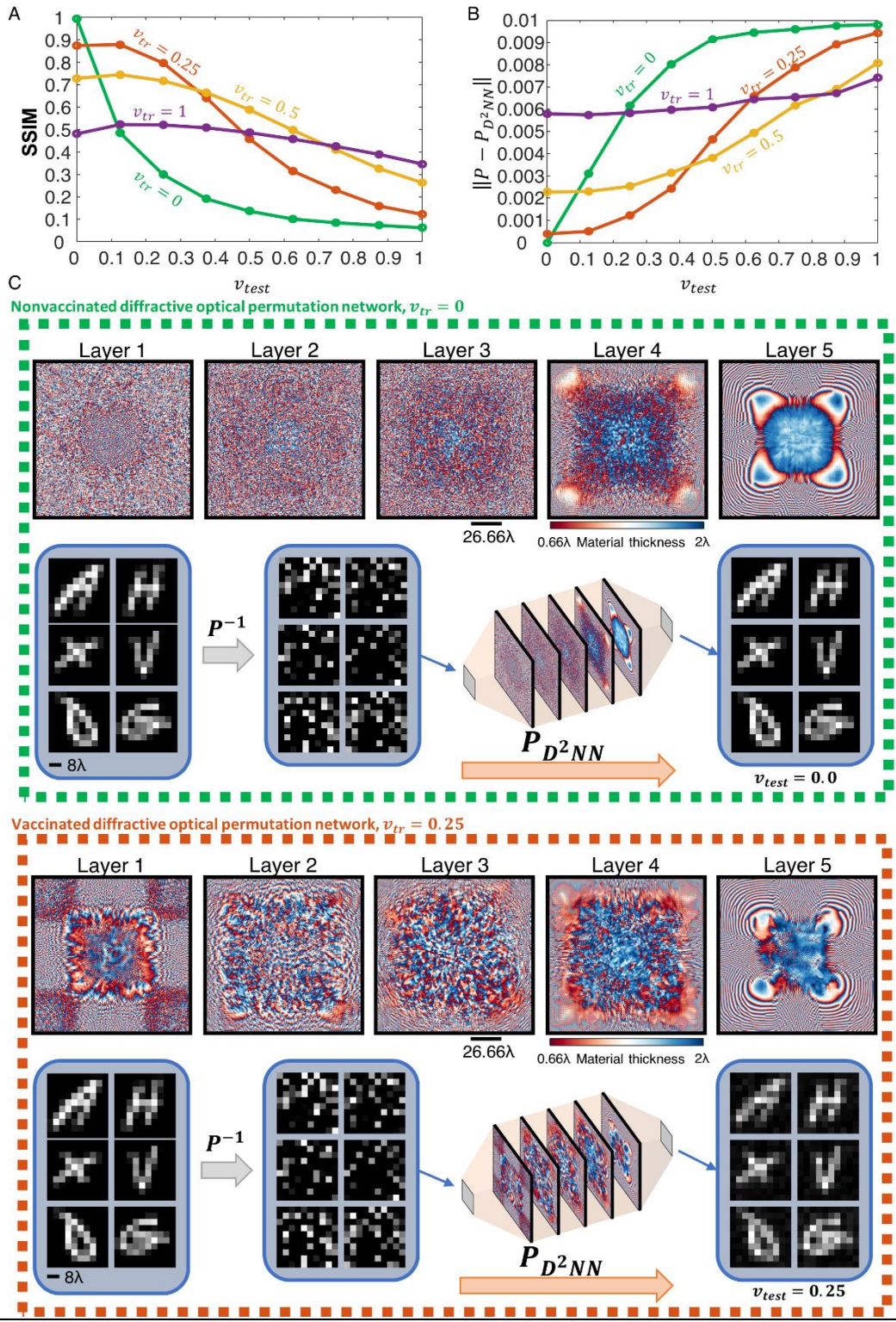

**Fig. 4: The sensitivity of the diffractive permutation networks against various levels of physical misalignments** A SSIM values achieved by 5-layer diffractive permutation networks with and without vaccination. B Transformation errors between the desired 100×100 permutation operation ($P$) and its optically synthesized counterpart ($P_{D^2NN}$) at different levels of misalignments denoted by $v_{test}$. C The layers of a nonvaccinated diffractive permutation network, i.e., $v_{tr} = 0$, along with the examples of EMNIST test images recovered optically through the diffractive permutation operation. D Same as C, except for a vaccinated diffractive permutation network based on $v_{tr} = 0.25$.

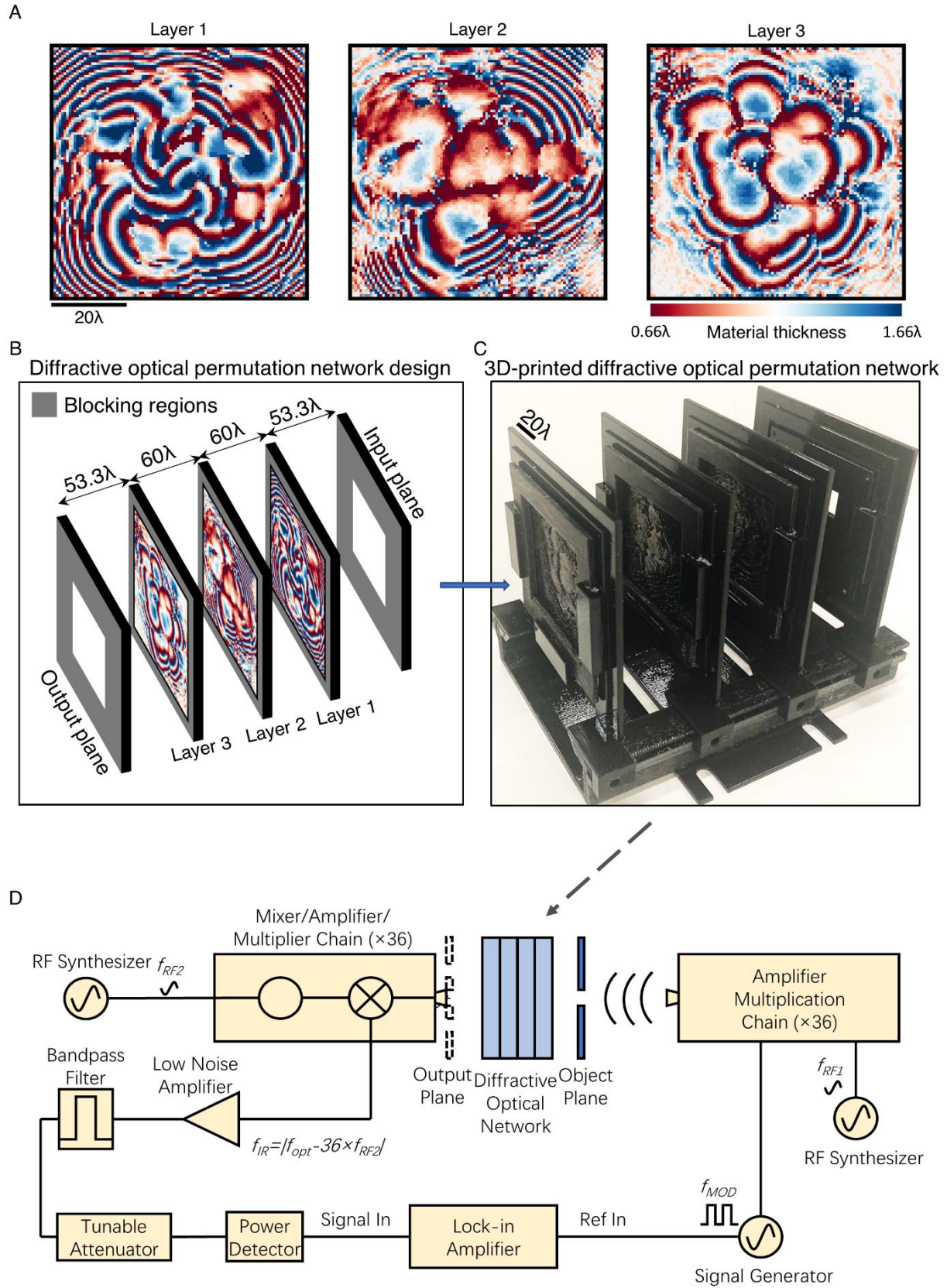

**Fig. 5: Experimental demonstration of a diffractive permutation network.** A The material thickness profiles of the diffractive surfaces of the fabricated diffractive permutation network. B The schematic of the experimental architecture illustrating the forward optical model of the diffractive permutation network. C 3D printed diffractive permutation network operating at THz part of the spectrum. D The schematic of our experimental system.

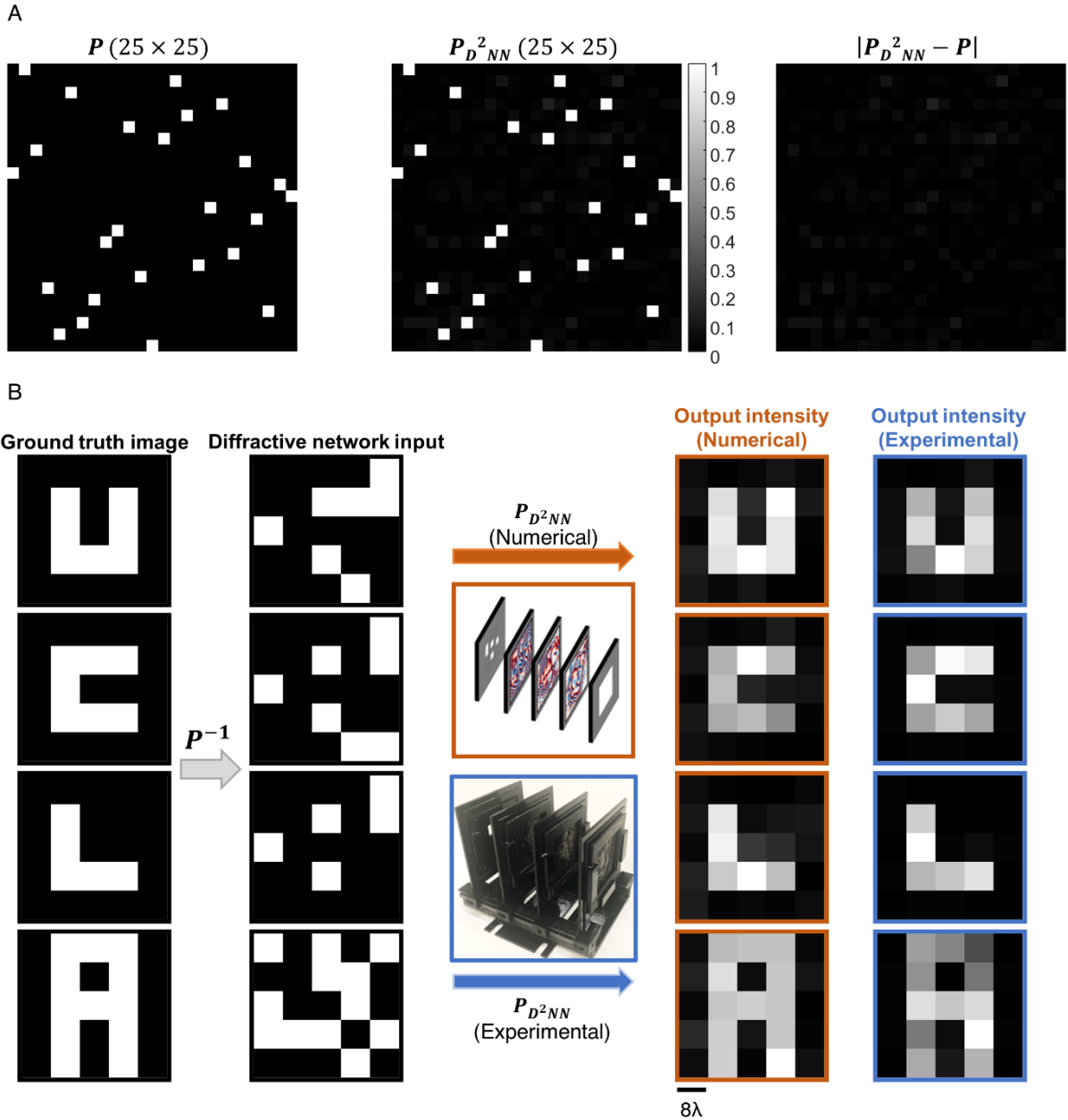

**Fig. 6: Experimental results.** A (left) The desired 25x25 permutation matrix, **P**, (middle) the optically realized permutation operation predicted by the numerical forward model, $\boldsymbol{P}_{D^2NN}$, and (right) the absolute error map between the two matrices. B Comparison between the numerically predicted and the experimentally measured output images for the task of recovering intensity patterns describing the letters 'U', 'C', 'L' and 'A'.